\documentclass[fleqn,twoside]{article}
\usepackage[headings]{espcrc2}
\usepackage[latin1]{inputenc}
\usepackage{alltt,amsmath,amssymb,colordvi,graphicx,pifont,feynarts}

\makeatletter
\def\={\penalty\@M-\hskip\z@skip}
\makeatother

\newcommand{\eg}{e.g.\ }
\newcommand{\ie}{i.e.\ }
\newcommand{\bM}{\mathbf{M}}
\newcommand{\cM}{\mathcal{M}}
\newcommand{\cA}{\mathcal{A}}
\newcommand{\ri}{\mathrm{i}}
\newcommand{\rd}{\mathrm{d}}
\newcommand{\re}{\mathrm{e}}
\newcommand{\limfunc}[1]{\mathop{\mathrm{#1}}}
\renewcommand{\Re}{\limfunc{Re}}
\renewcommand{\Im}{\limfunc{Im}}

\newcommand{\Code}[1]{\ensuremath{\text{\tt #1}}}
\newcommand{\Var}[1]{\ensuremath{{\it #1}}}

\newcommand{\eff}{\mathrm{eff}}
\newcommand{\aeff}{\ensuremath{\alpha_\eff}}

\newcommand{\BR}{\limfunc{BR}}

\newcommand\ML[1]{M_{\tilde L,#1}}
\newcommand\MR[1]{M_{\tilde R,#1}}

\newcommand\ru{\mathrm{u}}
\newcommand\rc{\mathrm{c}}
\newcommand\rt{\mathrm{t}}

\newenvironment{Tightemize}%
  {\leftmargini=10bp%
   \begin{itemize}%
   \itemsep=0pt plus .5pt%
   \parsep=0pt}%
  {\end{itemize}}

\newcommand\asat{\text{\ding{192}}}
\newcommand\twoloop{\text{\ding{193}}}
\newcommand\oneloop{\text{\ding{194}}}
\newcommand\order[1]{\ensuremath{\mathcal{O}(#1)}}

\runtitle{FeynHiggs 2.7}
\runauthor{T. Hahn et al.}

\begin{document}

\title{FeynHiggs 2.7}

\author{%
  T.~Hahn\address[MPI]{%
    Max-Planck-Institut f\"ur Physik,
    F\"ohringer Ring 6,
    D--80805 M\"unchen, Germany},
  S.~Heinemeyer\address{%
    Instituto de Fisica de Cantabria (CSIC-UC),
    Santander, Spain},
  W.~Hollik\addressmark[MPI],
  H.~Rzehak\address{%
    Institut f\"ur Theoretische Physik,
    Karlsruhe Institute for Technology,
    D--76128 Karlsruhe, Germany},
  G.~Weiglein\address{%
    DESY,
    Notkestr. 85,
    D--22607 Hamburg, Germany}}

\date{}

\begin{abstract}
We present the Version 2.7 of FeynHiggs, a program for computing MSSM 
Higgs-boson masses and related observables, such as mixing angles, 
branching ratios, and couplings, including state-of-the-art higher-order 
contributions.
\hfill\hbox{MPP-2010-78, KA-TP-21-2010, SFB/CPP-10-56, DESY 10-100}
\end{abstract}

\maketitle


\section{Complex Parameters in the MSSM Higgs Sector}

The Higgs sector of the Minimal Supersymmetric Standard Model with
complex parameters (cMSSM) consists of two Higgs doublets
\begin{align}
H_1 &= \begin{pmatrix}
       v_1 + \frac 1{\sqrt 2}(\phi_1 - \ri\chi_1) \\
       -\phi_1^-
       \end{pmatrix}, \\
H_2 &= \Magenta{\re^{\ri\xi}}
       \begin{pmatrix}
       \phi_2^+ \\
       v_2 + \frac 1{\sqrt 2}(\phi_2 + \ri\chi_2)
       \end{pmatrix}
\end{align}
which form the following Higgs potential:
\begin{equation}
\kern -2.5em
\begin{aligned}
V &= m_1^2\,H_1\bar H_1 + m_2^2\,H_2\bar H_2 - \\
&\quad (\Magenta{m_{12}^2}\,\varepsilon_{\alpha\beta}
       H_1^\alpha H_2^\beta + \mathrm{h.c.}) + \\
&\quad \frac{g_1^2 + g_2^2}{8}\,
       (H_1\bar H_1 - H_2\bar H_2)^2 +
       \frac{g_2^2}{2}\,|H_1\bar H_2|^2.
\end{aligned}
\end{equation}
The Higgs potential contains two complex phases \Magenta{$\xi$},
\Magenta{$\arg(m_{12}^2)$}.  The phase \Magenta{$\arg(m_{12}^2)$} can be
rotated away \cite{Peccei,MSSMcomplphasen} and, at tree level,
\Magenta{$\xi$} has to vanish in order to fulfill the minimum condition
of the Higgs potential, so there is no CP-violation at tree level and
the spectrum contains five states of definite CP-parity: $h$, $H$, $A$,
$H^\pm$.

CP-violating effects are induced by complex parameters that enter via 
loop corrections: the trilinear couplings $A_{t,b,\tau}$, the Higgsino 
mass parameter $\mu$, and the gaugino mass parameters $M_{1,2,3}$.
They yield $\Magenta{\hat\Sigma_{hA}}$, $\Magenta{\hat\Sigma_{HA}}\neq 
0$ and induce mixing between $h$, $H$, and $A$~\cite{mhiggsCPV}.  The 
Higgs mass matrix has the form
\begin{small}
\begin{equation}
\kern -2.8em
\bM^2 = \begin{pmatrix}
q^2 - m_h^2 + \hat\Sigma_{hh} &
        \hspace*{1em}\hat\Sigma_{hH} &
                \hspace*{1em}\Magenta{\hat\Sigma_{hA}} \\[.4ex]
\hat\Sigma_{Hh}\hspace*{1em} &
        \hspace*{-2em}
        q^2 - m_H^2 + \hat\Sigma_{HH}
        \hspace*{-2em} &
                \hspace*{1em}\Magenta{\hat\Sigma_{HA}} \\[.4ex]
\Magenta{\hat\Sigma_{Ah}}\hspace*{1em} &
        \Magenta{\hat\Sigma_{AH}}\hspace*{1em} &
                q^2 - m_A^2 + \hat\Sigma_{AA}
\end{pmatrix}
\notag
\end{equation}
\end{small}%
where $m_{h,H,A}$ denote the tree-level Higgs masses and the 
self-energies' momentum dependence has only been suppressed for compact 
notation.  It should be noted that in general $\bM^2$ is symmetric but 
not Hermitian.  In the approximation of vanishing external momentum 
($q^2 = 0$), one can obtain the higher-order corrected mass eigenstates 
via a unitary transformation from the tree-level states:
\begin{equation}
\label{def:U}
\begin{pmatrix}
  h_1 \\ h_2 \\ h_3
\end{pmatrix} = \begin{pmatrix}
  U_{11} & U_{12} & \Magenta{U_{13}} \\
  U_{21} & U_{22} & \Magenta{U_{23}} \\
  \Magenta{U_{31}} & \Magenta{U_{32}} & U_{33}
\end{pmatrix} \begin{pmatrix}
  h \\ H \\ A
\end{pmatrix}.
\end{equation}


\section{Higgs-boson Self-Energy Corrections in FeynHiggs}

\subsection{Higgs-boson Masses}

The following contributions to the mass matrix and the 
charged-Higgs-boson self-energy are taken into account:
\begin{gather}
\label{eq:massmatrix}
\kern -2.5em
\begin{pmatrix}
q^2 - m_h^2 + \hat\Sigma_{hh}^{\asat\twoloop\oneloop} &
        \hspace*{1.1em}\hat\Sigma_{hH}^{\asat\twoloop\oneloop} &
             \hspace*{1.2em}\Magenta{\hat\Sigma_{hA}^{\asat\oneloop}} \\[.4ex]
\hat\Sigma_{Hh}^{\asat\twoloop\oneloop}\hspace*{1.2em} &
        \hspace*{-2em}
        q^2 - m_H^2 + \hat\Sigma_{HH}^{\asat\twoloop\oneloop}
        \hspace*{-2em} &
             \hspace*{1.2em}\Magenta{\hat\Sigma_{HA}^{\asat\oneloop}} \\[.4ex]
\Magenta{\hat\Sigma_{Ah}^{\asat\oneloop}}\hspace*{1.2em} &
        \Magenta{\hat\Sigma_{AH}^{\asat\oneloop}}\hspace*{1.1em} &
                q^2 - m_A^2 + \hat\Sigma_{AA}^{\asat\oneloop}
\end{pmatrix} \notag \\
\hat\Sigma_{H^+H^-}^{\asat\oneloop}
\end{gather}

\begin{Tightemize}
\item[\asat]
Leading \order{\alpha_t\alpha_s} cMSSM two-loop corr.\ \cite{asatcplx}.

\item[\twoloop]
Leading \order{\alpha_t^2} and subleading 
$\mathcal{O}(\alpha_b\alpha_s$, $\alpha_t\alpha_b$, $\alpha_b^2)$ 
two-loop corrections evaluated in the real MSSM (rMSSM) 
\cite{atat,asab,atab}.  For the treatment of phases see 
Sect.~\ref{sect:tlcorr} below.

\item[\oneloop]
Full one-loop evaluation (all phases, $q^2$ dependence)
\cite{mhcMSSMlong} and leading non-minimal flavour-violating
(NMFV) corrections \cite{NMFV}. 
\end{Tightemize}
FeynHiggs \cite{feynhiggs,mhiggslong,mhiggsAEC,mhcMSSMlong} performs a
numerical search for the complex roots of $\det\bM^2(q^2)$ which are
denoted $\cM^2_{h_{1,2,3}}$.  We decompose $\cM^2 = M^2 - i M \Gamma$, 
where $M$ is the mass of the particle and $\Gamma$ its width, and then
define the loop\=corrected masses according to $M_{h_1} \leqslant 
M_{h_2} \leqslant M_{h_3}$.

The Higgs masses are thus determined as the real parts of the complex
poles of the propagator.  Complex contributions to the Higgs mass matrix
(from $\Im\hat\Sigma$) are taken into
account~\cite{mhcMSSMlong,lcws07FH}.


\subsection{Two-loop Corrections in the cMSSM}
\label{sect:tlcorr}

The full phase dependence is taken into account in the complete
one-loop~\cite{mhcMSSMlong} and \order{\alpha_t\alpha_s}
two-loop~\cite{asatcplx} contributions to the Higgs self-energies.  We
used the approximation of vanishing external momenta and vanishing
electroweak gauge couplings in the evaluation of all two-loop diagrams.

The \Code{tlCplxApprox} flag controls the treatment of phases in the 
part of the two-loop corrections known only in the rMSSM so far.  
The following values are possible:
\begin{Tightemize}
\item 0 = all corrections: \order{\alpha_t\alpha_s, \alpha_b\alpha_s,
  \alpha_t^2, \alpha_t\alpha_b, \alpha_b^2} in the rMSSM,
\item 1 = only the cMSSM \order{\alpha_t\alpha_s} corrections,
\item 2 = the cMSSM \order{\alpha_t\alpha_s} corrections combined with
  the remaining corrections in the rMSSM, truncated in the phases,
\item 3--6 = the cMSSM \order{\alpha_t\alpha_s} corrections combined 
  with the remaining corrections in the rMSSM, interpolated in the 
  phases.
\item New in 2.7 is the choice of interpolation variables:
  $A_t,A_b,M_3,\mu$ (3),
  $A_t,X_b,M_3,\mu$ (4),
  $X_t,A_b,M_3,\mu$ (5),
  $X_t,X_b,M_3,\mu$ (6).
\end{Tightemize}
FeynHiggs thus not only has the most precise evaluation of the Higgs 
masses in the cMSSM available to date (using the Feynman\=diagrammatic 
approach), but can also deliver an estimate of the uncertainties due 
to the rMSSM-only parts, by comparing the output for different values
of \Code{tlCplxApprox}.


\subsection{Improvements in the $\Delta_b$ Resummation}

Corrections to the bottom-Yukawa coupling are potentially large and 
important phenomenologically, in particular for large values of 
$\tan\beta$.

A resummation of the terms \order{\alpha_s^n \tan^n\beta} and 
\order{\alpha_t^n \tan^n\beta} can be performed with the help of 
$\Delta_b$ \cite{deltab1,deltab2}.

Some care has to be taken when applying the resummed $\Delta_b$ 
corrections in FeynHiggs because the $\Delta_b$ terms inserted into 
the one-loop Higgs self-energies have to be consistent with the 
two-loop corrections selected.  Whenever the corrections \twoloop\ in
Eq.~(\ref{eq:massmatrix}) are enabled (\Code{tlCplxApprox} $\neq$ 1), 
we also have to use the $\Delta_b$ of Ref.~\cite{asab}.  
The determination of $\Delta_b$ was improved in this direction in 
Version 2.7, avoiding incomplete higher-order corrections at 
\order{\alpha_b\alpha_s} and beyond.

In all other instances (\eg decays, production cross-sections), the
$\Delta_b$ of \cite{SchererDeltaB} is used in FeynHiggs 2.7, computed
self-consistently according to the procedure in \cite{BurasDeltaB}.


\section{Decay Rates and Production Cross-Sections}

\subsection{$h_i\to f_j \bar f_k$ at One-Loop Precision}

The Higgs decays to fermions, $h_i\to f_j\,\bar f_k$ are now
available at full one-loop precision \cite{WeigleinWilliams}.  
The real gluon (photon) which cancels the IR pole is treated fully 
inclusively \cite{BraatenLeveille}.

The routine \Code{FHCouplings} has an additional flag \Code{fast} to 
switch off the computation of the off-diagonal decays ($j\neq k$) for 
efficiency.  These decays hardly contribute to the total decay rate
and can safely be neglected.

The phenomenologically important resummed $\Delta_b$ corrections are 
still taken into account in $h_i\to b\bar b$, with the corresponding 
one-loop contribution subtracted to prevent double counting.


\subsection{Improvements in the $gg\to h$ Production Cross-Section}

For the Standard Model estimate, we use the NNLL prediction of
Ref.~\cite{Grazzini}.

The MSSM production cross-section is estimated in the effective-coupling
approximation \cite{HiggsXS} by multiplying pieces of the amplitude with 
the corresponding state-of-the-art amplitude correction factors as follows:
\begin{gather}
\cA^{\text{MSSM}}
= c_t^{\text{NLO}} c_t^{\text{NNLO}} \cA_t^{\text{MSSM,LO}} + \\
\notag
\qquad c_{b,r} \Re\cA_b^{\text{MSSM,LO}} +
      c_{b,i} \Im\cA_b^{\text{MSSM,LO}} + \\
\notag
\qquad c_{\tilde f} \cA_{\tilde f}^{\text{MSSM,LO}} +
      \cA_{\text{rest}}^{\text{MSSM,LO}}, \\[1ex]
\cA^{\text{SM}}
= c_t^{\text{NLO}} \cA_t^{\text{SM,LO}} + \\
\notag
\qquad c_{b,r} \Re\cA_b^{\text{SM,LO}} +
      c_{b,i} \Im\cA_b^{\text{SM,LO}} + \\
\notag
\qquad \cA_{\text{rest}}^{\text{MSSM,LO}}, \\[1ex]
\sigma^{\text{MSSM}} = \frac{|\cA^{\text{MSSM}}|^2}{|\cA^{\text{SM}}|^2}
  \sigma^{\text{SM,NLO}}.
\end{gather}
The factors $c_t^{\text{NLO}}$, $c_t^{\text{NNLO}}$, $c_{b,r}$, 
$c_{b,i}$ are taken from Ref.~\cite{kfactor} and by construction 
reproduce the top and bottom loops at NLO including the interference
term.  $c_{\tilde f}$ is taken from Ref.~\cite{kfactorSf}.
The LO cross-sections are parameterized with $m_b(m_b)$.

The $\sigma^{\text{SM,NLO}}$ is also taken from Ref.~\cite{kfactor} and
contains only SM top and bottom loops.


\section{Non-Minimal Flavour Violation}

In the NMFV MSSM, the sfermion flavours are allowed to mix with each 
other, \ie the mixing is 6$\times$6 rather than 2$\times$2:

\vskip -2ex

\begin{small}
$$
\begin{array}{c|c}
\text{NMFV} & \text{MFV} \\ \hline & \\[-1ex]
\tilde u_i = R^\ru_{ij} \begin{pmatrix}
  \tilde u_L \\
  \Green{\tilde c_L} \\
  \Blue{\tilde t_L} \\
  \tilde u_R \\
  \Green{\tilde c_R} \\
  \Blue{\tilde t_R}
\end{pmatrix}_{\!\!j} ~&
\begin{aligned}
\tilde u_i &= U^\ru_{ij} \begin{pmatrix}
  \tilde u_L \\
  \tilde u_R \\
\end{pmatrix}_{\!\!j} \\
\tilde c_i &= U^\rc_{ij} \begin{pmatrix}
  \Green{\tilde c_L} \\
  \Green{\tilde c_R} \\
\end{pmatrix}_{\!\!j} \\
\tilde t_i &= U^\rt_{ij} \begin{pmatrix}
  \Blue{\tilde t_L} \\
  \Blue{\tilde t_R}
\end{pmatrix}_{\!\!j}
\end{aligned}
\end{array}
$$
\end{small}%
and likewise for the sdown sector.

The mixing matrices $R^{\ru,\rd}$ diagonalize the mass matrices
$(M^2)^{\ru,\rd} = (M^2_{\text{MFV}})^{\ru,\rd} + \Delta^{\ru,\rd}$,
where

\begin{scriptsize}
\begin{equation*}
\kern -1.7em
\arraycolsep=1pt
M^2_{\text{MFV}} = \left(\begin{array}{ccc|ccc}
\ML{i}^2 & 0 & 0  &  m_i X_i^* & 0 & 0 \\
0 & \Green{\ML{j}^2} & 0  &  0 & \Green{m_j X_j^*} & 0 \\
0 & 0 & \Blue{\ML{k}^2}  &  0 & 0 & \Blue{m_k X_k^*} \\[.3ex]
\hline
m_i X_i & 0 & 0 &  \MR{i}^2 & 0 & 0 \\
0 & \Green{m_j X_j} & 0 &  0 & \Green{\MR{j}^2} & 0 \\
0 & 0 & \Blue{m_k X_k} &  0 & 0 & \Blue{\MR{k}^2}
\end{array}\right).
\end{equation*}
\end{scriptsize}%
The four sectors are clockwise labelled LL, LR, RR, RL with entries
\begin{small}
\begin{align*}
\ML{q}^2 &= M_{\tilde Q,q}^2 +
  m_q^2 + \cos 2\beta\,(T_3^q - Q_q s_W^2) m_Z^2\,, \\
\MR{q}^2 &= M_{\tilde U/\tilde D,q}^2 +
  m_q^2 + \cos 2\beta\,Q_q s_W^2 m_Z^2\,, \\
X_q &= A_q - \mu^*\tan^{-2 T_3^q}\beta\,.
\end{align*}
\end{small}%
Technical remark: FeynHiggs 2.7 keeps the MFV mixing matrices $U$
exactly `on top' of the NMFV matrices $R$, such that the MFV entries 
are automatically updated whenever the NMFV entries change and vice 
versa.

The most immediately notable effect comes from the LR(RL) sector,
as the $A^f_{ij}$ enter the couplings directly, \eg

\vskip -2ex

\noindent\begin{feynartspicture}(80,80)(1,1)
\FADiagram{}
\FAProp(20.,15.)(11.,10.)(0.,){ScalarDash}{-1}
\FALabel(15.2273,13.3749)[br]{$\tilde d_i$}
\FAProp(20.,5.)(11.,10.)(0.,){ScalarDash}{1}
\FALabel(15.2273,6.62506)[tr]{$\tilde d_j$}
\FAProp(0.,10.)(11.,10.)(0.,){ScalarDash}{0}
\FALabel(5.5,9.18)[t]{$A$}
\FAVert(11.,10.){0}
\end{feynartspicture}\raise 37bp\hbox{$\propto$}
\vskip -6ex
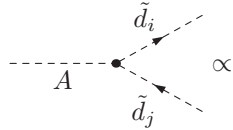
\begin{align*}
\sum_{g,g'} \Bigl[\, &
m_{d_{g'}} R_{i,g+3}^{\rd*} R_{j,g'}^{\rd}
  (\delta_{gg'} \mu + A_{g'g}^{d*} \tan\beta) - \\[-3ex] &
m_{d_g} R_{i,g}^{\rd*} R_{j,g'+3}^{\rd}
  (\delta_{gg'} \mu^* + A_{gg'}^{\rd} \tan\beta)
\,\Bigr]
\end{align*}
This enters the Higgs masses through the $A$ self-energy
and can lead to sizable effects.

The main constraints on the $\Delta^{\ru,\rd}$ come from low-energy 
observables.  Currently included in FeynHiggs are $b\to s\gamma$ and 
$\Delta M_s$, both at one-loop level including NMFV effects, with more 
to follow.


\section{Output of FeynHiggs 2.7}

We give a short overview of the output routines of the FeynHiggs 
library.

\medskip

\noindent\textbf{\Code{FHHiggsCorr}}
-- All Higgs-boson masses and mixings:
$M_{h_{1,2,3}}$, $M_{H^\pm}$, $\aeff$, \Code{UHiggs}, \Code{ZHiggs}.

\medskip

\noindent\textbf{\Code{FHUncertainties}}
-- Uncertainties of the masses and mixings.

\medskip

\noindent\textbf{\Code{FHCouplings}}
-- Couplings and Branching Ratios for the following decay channels:
\begin{align*}
h_{1,2,3} \to {}
& f_i\bar f_j, \gamma\gamma, ZZ^{(*)}, WW^{(*)}, gg,\\
& h_i Z^*, h_i h_j, H^+ H^-, \\
& \tilde f_i \tilde f_j,
  \tilde\chi_i^\pm \tilde\chi_j^\pm, \tilde\chi_i^0 \tilde\chi_j^0,
\\[1ex]
H^\pm \to {}
& f^{(*)}\bar f', h_i W^{\pm *},
  \tilde f \tilde f',
  \tilde\chi_i^0 \tilde\chi_j^\pm,
\\[1ex]
t\to {}
& W^+ b, H^+ b,
\end{align*}
plus the corresponding ones of an SM Higgs with mass $M_{h_i}$:
$h_{1,2,3}^{\text{SM}}\to f_i\bar f_j$, $\gamma\gamma$, $ZZ^{(*)}$, 
$WW^{(*)}$, $gg.$

\medskip

\noindent\textbf{\Code{FHHiggsProd}}
-- Higgs production-channel cross-sec\-tions
(SM total cross-sections multiplied with MSSM effective couplings, see
Ref.~\cite{Grazzini,HiggsXS,Kilgoreetal})

\begin{Tightemize}
\item
$gg\to h_i$ -- gluon fusion.

\item
$WW\to h_i$, $ZZ\to h_i$ -- gauge-boson fusion.

\item
$W\to W h_i$, $Z\to Z h_i$ -- Higgs-strahlung.

\item
$b\bar b\to b\bar b h_i$ -- bottom Yukawa process.

\item
$b\bar b\to b\bar b h_i$ -- ditto, one $b$ tagged.

\item
$t\bar t\to t\bar t h_i$ -- top Yukawa process.

\item
$\tilde t \bar{\tilde t}\to \tilde t\bar{\tilde t} h_i$
-- stop Yukawa process.
\end{Tightemize}

\medskip

\noindent\textbf{\Code{FHConstraints}}
-- Electroweak precision observables, see \eg Ref.~\cite{PomssmRep} for
details:

\begin{Tightemize}
\item
$\Delta\rho$ at \order{\alpha, \alpha\alpha_s}, including NMFV effects.

\item
$M_W$, $\sin^2\theta_\eff$ via SM formula + $\Delta\rho$.

\item
$\BR(b\to s\gamma)$ and $\Delta M_s$ at one-loop level including
NMFV effects \cite{bsgNMFV}.

\item
$(g_\mu - 2)_{\mathrm{SUSY}}$
including full one- and leading/ subleading two-loop SUSY 
corrections.

\item
EDMs of electron (Th), neutron, Hg.
\end{Tightemize}


\section{Download and Build}

\begin{Tightemize}
\item
Get the tar file from feynhiggs.de, unpack and configure:
\begin{verbatim}
tar xfz FeynHiggs-2.7.0.tar.gz
cd FeynHiggs-2.7.0
./configure
\end{verbatim}

\item
``\Code{make}'' builds the Fortran/C++ part only.
``\Code{make all}'' builds also the Mathematica part.

\item
``\Code{make install}'' installs the package.

\item
``\Code{make clean}'' removes unnecessary files.
\end{Tightemize}


\section{Usage}

FeynHiggs has four modes of operation: Library mode, Command-line mode, 
Web mode, and Mathematica mode.


\subsection{Library Mode}

The FeynHiggs library \Code{libFH.a} is a static Fortran 77 library.
Its global symbols are prefixed with a unique identifier to minimize
symbol collisions.  The library contains only subroutines (no
functions), so that no include files are needed (except for the
couplings) and the invocation from C/C++ is hassle-free.  Detailed
debugging output can be turned on at run time.  All routines are
described in detail in the API guide and on man-pages.


\subsection{Command-Line Mode}
\label{sec:commandline}

The user submits a text file, such as
\begin{center}
\begin{small}
\begin{minipage}[t]{.5\hsize}
\begin{verbatim}
MT           172.6
MSusy        500
MA0          200
Abs(M_2)     200
Abs(MUE)     1000
TB           5
Abs(Xt)      1000
Abs(M_3)     800
\end{verbatim}
\end{minipage}
\end{small}
\end{center}
to the FeynHiggs executable with a command like
\begin{alltt}
FeynHiggs \Var{file} [\Var{flags}]
\end{alltt}
where the \Var{flags} are optional.  The output is a human-readable 
version of the results.  The \Code{table} utility converts the output 
to machine-readable format, for example
\begin{alltt}
FeynHiggs \Var{file} [\Var{flags}] | table TB Mh0 > \Var{out}
\end{alltt}
Loops over parameter values (parameter scans) are possible:
\begin{Tightemize}
\item
\Code{MA0 200 350 50}, linear: 200, 250, 300, 350,

\item
\Code{TB 5 40 *2}, logarithmic: 5, 10, 20, 40,

\item
\Code{TB 5 50 /4}, number of steps: 5, 20, 35, 50.
\end{Tightemize}


\subsubsection{Command-Line Mode Scripted}

If a ``\Code{-}'' is specified as file name in command-line mode, 
FeynHiggs reads the input from stdin.  This allows to script FeynHiggs 
sessions, such that parameters, flags, and possibly environment 
variables are preserved in a compact shell script.  For example:

\begin{scriptsize}
\begin{verbatim}
#! /bin/sh

make || exit 1

FHDEBUG=2 ./build/FeynHiggs - ${1:-400302103} << END
MT           173.1
MSusy        3000
MA0          1000
Abs(M_2)     2500
Abs(MUE)     2000
TB           5 
Abs(Xt)      1000
Abs(M_3)     2000
END
\end{verbatim}   
\end{scriptsize}

\Code{make || exit 1} updates the FeynHiggs executable and terminates 
with an error if any problem occurred in the build.

The prefixed \Code{FHDEBUG=2} sets an environment variable, here to 
increase the debugging level.

There are default flags, overridable through command-line argument 1
(\verb=${1:-400302103}=).

The actual parameters are fed to FeynHiggs up to the \Code{END}
marker.  The identifier \Code{END} can be chosen freely but must match
exactly the one at the bottom of the parameter list.


\subsection{SUSY Les Houches Accord Format}

The \Code{FeynHiggs} executable can also process files in SUSY Les
Houches Accord 2 (SLHA2) format \cite{SLHA2}.  It uses the SLHA Library
\cite{SLHALib}.  Processing of SLHA2 files can also be done in Library
Mode with the subroutine \Code{FHSetSLHA}.

FeynHiggs in fact tries to read each file in SLHA format first and 
falls back to its native format if that fails.


\subsection{Web Mode}

The FeynHiggs User Control Center (FHUCC) is online at feynhiggs.de/fhucc.  
It is a Web interface for the command-line frontend.  The user gets the 
results together with the input file for the command-line frontend.


\subsection{Mathematica Mode}

A more powerful interactive environment is provided by the Mathematica
interface of FeynHiggs.  The MathLink executable \Code{MFeynHiggs} must
first be loaded with
\begin{verbatim}
   Install["MFeynHiggs"]
\end{verbatim}
and makes all FeynHiggs routines available as Mathematica functions.  In
combination with the arsenal of standard Mathematica functions such as
\Code{ContourPlot} and \Code{Manipulate}, even sophisticated analyses
can be carried out easily.


\section{Summary: Main New Features}

Version 2.7 of FeynHiggs includes the following new features:
\begin{Tightemize}
\item
Inclusion of the full cMSSM two-loop \order{\alpha_t\alpha_s}
corrections in highly optimized form.  Extended interpolation
options for rMSSM parts.

\item
Inclusion of the full one-loop corrections to the
$h_i\to f_j \bar f_k$ decays.

\item
Inclusion of NMFV corrections in the Higgs self-energies, decays,
and low-energy constraints.

\item
Improvements in the $\Delta_b$ resummation.

\item 
Total Higgs production cross-sections in effective coupling
approximation, with particular improvements in the $ggh$ channel.
\end{Tightemize}


\begin{flushleft}

\end{flushleft}

\end{document}